\documentclass[fleqn,usenatbib]{mnras}

\usepackage[T1]{fontenc}
\usepackage{ae,aecompl}
\usepackage{changes}
\usepackage{graphicx}	
\usepackage{amsmath}	
\usepackage{amssymb}	

\title[]{Testing the warmness of dark matter}

\author[S. Kumar, R.C. Nunes and S.K. Yadav]{
Suresh Kumar,$^{1}$\thanks{suresh.kumar@pilani.bits-pilani.ac.in}
Rafael C. Nunes,$^{2}$
Santosh Kumar Yadav$^{1}$
\\
$^{1}$Department of Mathematics, BITS Pilani, Pilani Campus, Rajasthan-333031, India\\
$^{2}$Divis\~ao de Astrof\'isica, Instituto Nacional de Pesquisas Espaciais, Avenida dos Astronautas 1758, S\~ao Jos\'e dos Campos, 12227-010, SP, Brazil\\
$\;$
}

\date{Accepted XXX. Received YYY; in original form ZZZ}

\pubyear{2019}

\begin{document}
\label{firstpage}
\pagerange{\pageref{firstpage}--\pageref{lastpage}}
\maketitle
\begin{abstract}
Dark matter (DM) as a pressureless perfect fluid provides a good fit of the standard $\Lambda$CDM model to the astrophysical and cosmological data. In this paper, we investigate two extended properties of DM: a possible time dependence of the equation of state of DM via Chevallier-Polarski-Linder parametrization,   $w_{\rm dm} = w_{\rm dm 0} + w_{\rm dm 1}(1-a)$, and the constant non-null sound speed $\hat{c}^2_{\rm s,dm}$. We analyze these DM properties on top of the base $\Lambda$CDM model by using the data from Planck cosmic microwave background (CMB) temperature and polarization anisotropy, baryonic acoustic oscillations (BAO) and the local value of the Hubble constant from the Hubble Space Telescope (HST). We find new and robust constraints on the extended free parameters of DM. The most tight constraints  are imposed by CMB + BAO data where the three parameters $w_{\rm dm0}$, $w_{\rm dm1}$  and $\hat{c}^2_{\rm s,dm}$ are respectively constrained to be less than $1.43\times 10^{-3}$, $1.44\times 10^{-3}$ and $1.79\times 10^{-6}$ at 95\% CL. All the extended parameters of DM show consistency with zero at 95\% CL, indicating no evidence beyond the CDM paradigm. We notice that the extended properties of DM significantly affect several parameters of the base $\Lambda$CDM model. In particular, in all the analyses performed here, we find significantly larger mean values of $H_0$ and lower mean values of $\sigma_8$ in comparison to the base $\Lambda$CDM model. Thus, the well-known $H_0$ and $\sigma_8$ tensions might be reconciled in the presence of extended DM parameters within the $\Lambda$CDM framework. Also, we estimate the warmness of DM particles as well as its mass scale, and find a lower bound: $\sim$ 500 eV from our analyses. 

\end{abstract}



\section{Introduction}
\label{sec:intro}
Dark matter (DM) is a mysterious matter component of the Universe, and is expected to account for one-fourth of the energy budget of the Universe today. Over the years, many attempts have been made via direct and indirect searches to detect DM particle(s) but nothing convincing and conclusive is found so far. However, there are many physically motivated candidates of DM \citep{bertone2005particle}, which are assumed to behave as a pressureless perfect fluid usually modeled as ``Cold Dark Matter" (CDM). In the standard $\Lambda$CDM model, the major component ``cosmological constant" ($\Lambda$) is  associated with dark energy (DE) fluid whereas the CDM is considered as pressureless, non-interacting (except gravitationally) perfect fluid having zero equation of state (EoS) parameter as well as zero sound speed and zero viscosity. The consideration of CDM in the standard model leads to many small scale problems. For instance, the observed halo properties differ from the predictions of the standard $\Lambda$CDM model which might be an indication of DM being more complex than simply CDM. Many observed halo density profiles have cores in their centers rather than cusps \citep{moore1994evidence},
 and some have substructures \citep{jee2014hubble} that are at odds with the standard
$\Lambda$CDM simulations. Also, the low observed mass function of small halos seems to be in a serious disagreement with the results from the standard $\Lambda$CDM simulations \citep{boylan2011too, papastergis2015there}.

The proposed candidate, which may alleviate many of small scale issues, is the warm DM (WDM), and it is not distinguishable from CDM on larger scales. Therefore, the investigation of the precise nature of DM is important and worthwhile in modern cosmology. Several attempts have been done in the literature to understand the properties and precise nature of DM  by investigating its generalized or extended properties. In \cite{de2011warm}, it is claimed that the warmness of DM can successfully reproduce the astronomical observations from small to large scales. To test the warmness of DM, \cite{muller2005cosmological} has investigated the EoS parameter of DM using cosmic microwave background (CMB), Supernovae type Ia (SNe Ia), and large scale structure (LSS) data with zero adiabatic sound speed and no entropy production. \cite{faber2006combining} and  \cite{serra2011measuring} have constrained the EoS of DM by combining kinematic and gravitational lensing data. The warmness of DM has also been investigated in interacting and non-interacting scenarios of DM and DE  by assuming constant EoS of both DM and DE  in \cite{avelino2012testing} and  \cite{cruz2013interacting}.  Many models of DM and DE have been constrained by assuming constant as well as variable EoS parameter of DM and DE in \cite{wei2013cosmological}, where the authors have found that warmness of DM is not favored over coldness in the light of data from CMB, SNe Ia and baryonic acoustic oscillations (BAO). \cite{kumar2014observational} have investigated the fluid perspective of DM  and DE via variable EoS of both, and found no significant deviation from the CDM scenario, but obtained tighter constraints on  EoS of DM in comparison to previous similar studies \citep{calabrese2009cosmological, xu2013equation}.

In most of the above-mentioned works, the authors have focused only on constraining the EoS parameter of DM by considering it, either a constant or time-varying in different mathematical forms. The other generalized properties, like, sound speed and viscosity have been considered to be zero, either for simplicity or to avoid a large number of model parameters. 
Recently, an extensive investigation of the generalized properties of DM: EoS parameter, sound speed and viscosity (initially proposed in \cite{hu1998structure}), are discussed in detail by \cite{kopp2016extensive}. Next, \cite{thomas2016constraining} have found strong observational constraints on generalized DM parameters with recent observational datasets. 
Similar constraints on generalized DM properties have also been found by \cite{kunz2016constraints}. 
 Most recently, generalized DM properties have been investigated by \cite{tutusaus2018generalized} to reconcile the tension between Planck CMB and weak lensing observations. \cite{kopp2018dark} have tested the inverse cosmic volume law for DM by allowing its EoS to vary independently in eight redshift bins from $z = 10^5$ to $z = 0$ by using the latest observational data, and found no evidence for non-zero EoS parameter in any of the eight redshift bins.

In the recent past, it has been reported that Planck CMB observations and LSS observations are not in agreement with each other \citep{macaulay2013lower, battye2015tension, maccrann2015cosmic, lin2017cosmological}. The well-known and widely discussed discrepancies are in the measurements of Hubble constant $H_0$ and amplitude of present matter density $\sigma_8$, commonly known as $H_0-\sigma_8$ tensions. In recent years, many physical mechanisms have been investigated for alleviating these tensions with different perspectives \citep{bernal2016trouble,kumar2016probing,kumar2017echo,di2018vacuum,yang2018interacting,bringmann2018converting,kumar2018cosmological,yang2018tale,poulin2018implications,nunes2018structure,yang2019observational, feeney2019prospects,kumar2019dark}.

In the light of above discussion, in this paper, we are motivated to place robust and accurate constraints on some extended properties of DM such as its EoS parameter and sound speed, which are helpful to characterize the physical nature of DM.  Recently, the generalized DM parameters have been constrained by \cite{thomas2016constraining}, where all the parameters are taken as  constants. But, there is no reason for the EoS of DM to be a constant, it could be a time-varying as well. In this work,  we consider a time-dependent EoS and a constant sound speed of DM whereas viscosity is taken as zero (to avoid large number of parameters in the model). The role of DE is played by the cosmological constant. We use the recent cosmological observations including the data from CMB, BAO and Hubble Space Telescope (HST) to constrain the model parameters. In addition, we have also presented observational constraints on standard $\Lambda$CDM model with all the data sets under consideration, for comparison purpose. The paper is organized as follows: In the next section, the cosmological model with the extended DM properties is presented. Section \ref{data} presents the methodology and the data sets, which are used to constrain the free parameters of the considered model. In Section \ref{results}, the observational constraints are derived, and the results are discussed in detail. The last section carries the concluding remarks of this study.  

 \section{Model with extended properties of dark matter}
\label{the_model}
We consider Friedmann-Lemaitre-Robertson-Walker (FLRW) Universe, where the background expansion is governed by the so-called Friedmann equations (in the units $c=1$):
\begin{align}\label{model1}
3 H^2 =  8 \pi G \sum_{i}\rho_{i},\\ 
2 \frac{\text{d}H}{\text{d}t} + 3 H^2 = - 8 \pi G\sum_{i} P_i.
\end{align}
Here, $H=\frac{1}{a}\frac{da}{dt}$ is the Hubble parameter with $a$ being the scale factor of the Universe; $t$ is the cosmic time, and $G$ is the Newton's gravitational constant. Further, $\rho_i$ and $P_i$ are the energy density and pressure of the $i$th species, where the label $i$ runs over the components $i = \gamma, \nu, b, 
{\rm dm}, \Lambda $, representing photons, neutrinos, baryons, DM and cosmological constant, respectively. In what follows, a subindex
0 attached to any parameter denotes the value of the parameter at the present time.

In this work, we relax the condition that entire DM is purely a pressureless, non-relativistic component. For this, we assume, in principle, that EoS of the DM has a temporal dependence through the cosmic evolution. In order to quantify that we choose the functional form of the Chevallier-Polarski-Linder (CPL) parametrization \citep{chevallier2001accelerating, linder2003exploring} for the EoS of the DM, given by
\begin{align}\label{CPL}
w_{\rm dm}(a) = w_{\rm dm 0} + w_{\rm dm 1}(1-a),
\end{align}
where $w_{\rm dm0}$ and $w_{\rm dm1}$ are free parameters (constants) to be fixed by observations. For $w_{\rm dm0} = w_{\rm dm1} = 0 $, we recover $w_{\rm dm} = 0$, the EoS parameter of CDM. We assume that DM is described by a perfect fluid, and as usual quantified by the energy-momentum tensor with density $\rho$ and isotropic pressure $p$: $T_{\mu \nu} = (\rho + p) u_{\mu}  u_{\nu} + p g_{\mu \nu}$, where we have disregarded possible anisotropic stress tensor contribution. It is well known that anisotropic stress vanishes for perfect fluids or minimally coupled scalar fields. Taking $p_{\rm dm} = w_{\rm dm} \rho_{\rm dm}$ and the conservation law $\nabla_{\mu} T^{\mu \nu} = 0$, we have
\begin{align}
\frac{{\rm d\rho}_{\rm dm}}{\rm{d}t} + 3H[1 + w_{\rm dm}(a)]\rho_{\rm dm} = 0.
\end{align}

In the conformal Newtonian gauge, the perturbed FLRW metric takes the form
\begin{eqnarray}
\begin{aligned}
 \text{d}s^2  = a^2(\tau) \Big[-(1+2\psi) \text{d}\tau^2 + (1 - 2\phi)\text{d}\vec{r}^2 \Big],
\end{aligned}
\end{eqnarray}
where $\phi$ and $\psi$ are the metric potentials and $\vec{r}$ represents the three spatial coordinates. 
 In the Fourier space, the first order perturbed part of the conserved stress-energy momentum tensor, i.e., $ \delta T^{\mu \nu}_{; \nu} = 0$, leads to the following continuity and Euler equations \citep{ma1995cosmological} for DM:
\begin{eqnarray}\label{continiuty}
\begin{aligned}
 \dot{\delta}_{\rm dm}=-(1+w_{\rm dm})(\theta_{\rm dm}- 3 \dot{\phi}) - 3 \mathcal{H}\left(\frac{\delta p_{\rm dm}}{\delta \rho_{\rm dm}} - w \right) \delta_{\rm dm}, 
\end{aligned}
\end{eqnarray}

\begin{eqnarray}\label{Euler}
\begin{aligned}
\dot{\theta}_{\rm dm}  =  -(1-3 w_{\rm dm}) \mathcal{H} \theta_{\rm dm}  - \frac{\dot{w}_{\rm dm}}{1+w_{\rm dm}} \theta_{\rm dm}  + k^2\psi \\
+ \frac{\delta p_{\rm dm}}{\delta \rho_{\rm dm}} k^2 \frac{\delta_{\rm dm}}{1+w_{\rm dm}}. 
\end{aligned}
\end{eqnarray}
Here, an over dot stands for the conformal time derivative, $\mathcal{H}$ is the conformal Hubble parameter, and $k$ is magnitude of the wavevector $\vec{k}$. Further, $\delta_{\rm dm} = \delta \rho_{\rm dm} / \rho_{\rm dm}$ and $(\rho_{\rm dm} + p_{\rm dm})\theta_{\rm dm} = i k^j \delta T^{0}_{j}$  are the relative density and velocity perturbations, respectively, associated with the DM fluid. 
In a random frame, the quantity $\delta p_{\rm dm}/\delta \rho_{\rm dm}$ can be expressed as \citep{de2010measuring}, 
\begin{eqnarray}
\begin{aligned}
 \rho^{-1}_{\rm } \delta p_{\rm dm} =  \delta_{\rm dm} \hat{c}^2_{\rm s,dm} + 3 \mathcal{H} (1+w_{\rm dm}) (\hat{c}^2_{\rm s,dm} - c^2_{\rm a,dm})\frac{\theta_{\rm dm}}{k^2},
\end{aligned}
\end{eqnarray}
where $\hat{c}_{\rm s,dm}$ represents the sound speed of DM in the rest frame, and $c^2_{\rm a,dm}$ denotes the adiabatic sound speed squared, defined as

\begin{equation}
c^2_{\rm a,dm}=\frac{\dot{p}_{\rm dm}}{\dot{\rho}_{\rm dm}}=w_{\rm dm}-\frac{\dot{w}_{\rm dm}}{3\mathcal{H}(1+w_{\rm dm})}.
\end{equation}
The above two equations allow to recast (\ref{continiuty}) and (\ref{Euler}) as follows:

\begin{eqnarray}
\begin{aligned}
 \dot{\delta}_{\rm dm}= -(1+w_{\rm dm}) \left(\theta_{\rm dm} - 3 \dot{\phi} \right)   
 -3 \mathcal{H} \delta_{\rm dm} (\hat{c}^2_{\rm s,dm} - w_{\rm dm}) \\ - 9 (1+w_{\rm dm})(\hat{c}^2_{\rm s,dm} - c^2_{\rm a,dm})\mathcal{H}^2 \frac{\theta_{\rm dm}}{k^2},
\end{aligned}
\end{eqnarray}

\begin{eqnarray}
\begin{aligned}
\dot{\theta}_{\rm dm}&=&-(1-3 \hat{c}^2_{\rm s,dm}) \mathcal{H} \theta_{\rm dm}  + \frac{\hat{c}^2_{\rm s,dm}}{1+w_{\rm dm}}k^2 \delta_{\rm dm} + k^2\psi. 
\end{aligned}
\end{eqnarray}
 
 The sound speed of DM describes its micro-scale properties. Here, we consider $\hat{c}^2_{\rm s,dm}$  as a constant model parameter to be fixed by the observations. A significant deviation of sound speed from zero in light of the cosmological observations can be interpreted as a possible evidence for DM to be something different from the simple CDM. 
 
 Considering the background and perturbation dynamics presented above, in the next sections, we explore the full parameter space of the cosmological scenario provided by $\Lambda$CDM plus extended DM properties ($w_{\rm dm0}, \, w_{\rm dm1}, \, \hat{c}^2_{\rm s,dm}$). We name it $\Lambda$WDM model. The baseline free parameters set of this model is, therefore:

\begin{equation*}
\label{baseline1}
\begin{aligned}
\mathcal{P}_{\Lambda \rm WDM}= \Big\{ \omega_{\rm b}, \, \omega_{\rm dm}, \, \theta_s, \,  A_s, \, n_s, \, \tau_{\rm reio}, 
\,   w_{\rm dm0}, \, w_{\rm dm1}, \, \hat{c}^2_{\rm s,dm} \Big\},
\end{aligned}
\end{equation*}
where the first six parameters are the baseline parameters of the $\Lambda$CDM model, namely: $\omega_{\rm b}$ and $\omega_{\rm dm}$ are respectively the dimensionless densities of baryons and DM; $\theta_s$ is the ratio of the sound horizon to the angular diameter distance at decoupling; $A_s$ and $n_s$ are respectively the amplitude and spectral index of the primordial curvature perturbations, and $\tau_{\rm reio}$ is the optical depth to reionization.
\section{data sets and  methodology}
\label{data}

To constrain the free parameters of the $\Lambda$WDM model, we use the recent observational data sets described below.\\

\noindent\textbf{CMB}: CMB temperature and polarization data from  Planck-2015 \citep{ade2016planck},  comprised of likelihoods of low-$l$ temperature and polarization at $l \leq 29$, temperature (TT) at $l \geq 30$, cross correlation of temperature and  polarization (TE) and polarization (EE) power spectra. We also include Planck-2015 CMB lensing power spectrum likelihood \citep{ade2016planckGL}.\\

\noindent\textbf{BAO}: Four probes of baryon acoustic oscillations distance measurements to break the parameter degeneracy from other observations. These four measurements include the  Six  Degree  Field  Galaxy  Survey  (6dFGS) at redshift $z_{\rm eff} = 0.106$ \citep{beutler20116df}, 
the  Main  Galaxy  Sample  of  Data  Release 7  of  Sloan  Digital  Sky  Survey  (SDSS-MGS) at redshift $z_{\rm eff} = 0.15$ \citep{ross2015clustering}, 
the  LOWZ  and  CMASS  galaxy  samples  of Data Release 11 (DR11) of   the Baryon  Oscillation  Spectroscopic  Survey  (BOSS) LOWZ  and  BOSS-CMASS at redshifts $z_{\rm eff} = 0.32$ and $z_{\rm eff} = 0.57$, respectively \citep{anderson2014clustering}.  These data are summarized in \cite{nunes2016new}.\\

\noindent\textbf{HST}: Recently measured local
value of Hubble constant, $H_0=73.24 \pm 1.74$  km s${}^{-1}$ Mpc${}^{-1}$ by Hubble Space Telescope (HST), as reported in \cite{riess20162}.\\

\begin{table}
\caption{Uniform priors on the free parameters of the $\Lambda$WDM model.} \label{tab:priors}
\begin{center}
\begin{tabular}{|l|l|}
\hline
Parameter & Prior\\
\hline
$100 \omega_{\rm b}$ & [0.8, 2.4]\\
$\omega_{\rm dm}$ & [0.01, 0.99] \\
$100\theta_s$ & [0.5, 2.0] \\
$\ln[10^{10}A_{s }]$ & [2.7, 4.0]\\
$n_s$ & [0.9, 1.1] \\
$\tau_{\rm reio}$  & [0.01,  0.8] \\
$w_{\rm dm0}$  & [0, 0.1] \\
$w_{\rm dm1}$ &  [0, 0.1] \\
$\hat{c}^2_{\rm s,dm}$ & [0, 0.1]\\
\hline
\end{tabular}
\end{center}
\end{table}




\begin{table*}
\caption{\label{Table_M2} {Constraints on the free parameters and some derived parameters of $\Lambda$WDM model for four data combinations. 
The upper and lower values over the mean value of each parameter denote 68\% CL and 95\% CL errors. The parameter $H_{\rm 0}$ is measured in the units of km s${}^{-1}$ Mpc${}^{-1}$. The entries in blue color represent the constraints on the corresponding $\Lambda$CDM parameters.}}
\resizebox{\textwidth}{!}{%
\begin{tabular} { |l| l| l| l| l|  }  \hline \hline 
 Parameter &  CMB     & CMB + BAO        & CMB + HST       & CMB + BAO + HST     \\ 
\hline
$10^{2}\omega_{\rm b }$ &  $2.216^{+0.016+0.033}_{-0.016 -0.031} $ &      $2.217^{+ 0.015+0.031}_{-0.015-0.029}$ &  $2.218^{+ 0.016+0.032}_{-0.016-0.031}  $  & $2.218^{+0.016+0.032}_{-0.016-0.031}  $  \\[1ex]

& \textcolor{blue}{$2.226^{+0.015+0.030}_{-0.015 -0.029}$}    &  \textcolor{blue}{$2.235^{+ 0.014+0.027}_{-0.014-0.027}$} &    \textcolor{blue}{$2.238^{+0.015+0.030}_{-0.015 -0.029}$} &    \textcolor{blue}{$2.243^{+0.013+0.026}_{-0.013-0.026}  $} \\[1ex]
 \hline
 $\omega_{\rm dm }$  & $0.1156^{+0.0029+0.0047}_{-0.0021-0.0051}$ &  $0.1173^{+0.0011+0.0023}_{-0.0011-0.0022}  $ & $0.1173^{+0.0020+0.0038}_{-0.0020-0.0038}$  &  $0.1164^{+0.0011+0.0021}_{-0.0011-0.0021}   $   \\[1ex]
 
         & \textcolor{blue}{$0.1193^{+0.0014+0.0029}_{-0.0014-0.0028}$}   &  \textcolor{blue}{$0.1181^{+0.0010+0.0020}_{-0.0010-0.0020}  $}   & \textcolor{blue}{$0.1179^{+0.0013+0.0027}_{-0.0013-0.0025}$}  & \textcolor{blue}{$0.1173^{+0.0010+0.0020}_{-0.0010-0.0020}   $} \\[1ex]
\hline

$100 \theta_{s } $ &  $1.04166^{ +0.00031+0.00059}_{-0.00031-0.00062} $ & $1.04171^{+0.00031+0.00062}_{-0.00031-0.00059}$  &   $1.04166^{ +0.00032+0.00063}_{-0.00032-0.00065} $ & $1.04172^{+0.00032+0.00063}_{-0.00032-0.00061}$   \\[1ex]

&  \textcolor{blue}{$1.04185^{ +0.00029+0.00057}_{-0.00029-0.00056}$}   & \textcolor{blue}{$1.04197^{+0.00028+0.00055}_{-0.00028-0.00056}$} &  \textcolor{blue}{$1.04197^{ +0.00029+0.00057}_{-0.00029-0.00056}$}   &  \textcolor{blue}{$1.04204^{+0.00029+0.00055}_{-0.00029-0.00056}$}   \\[1ex]
\hline
$\ln10^{10}A_{s }$ &  $3.089^{+ 0.027+0.058}_{-0.030-0.053} $    & $3.082^{+0.025+0.047}_{-0.025-0.049}   $ &  $3.100^{+ 0.027+0.051}_{-0.027-0.052} $    &   $3.084^{+0.026+0.051}_{-0.026-0.052}$ \\[1ex]

&  \textcolor{blue}{$3.065^{+ 0.025+0.048}_{-0.025-0.050}  $}   & \textcolor{blue}{$ 3.077^{+0.023+0.044}_{-0.023-0.045}   $} &  \textcolor{blue}{$3.079^{+ 0.025+0.047}_{-0.025-0.049}  $}  & \textcolor{blue}{$3.087^{+0.022+0.045}_{-0.022-0.043}  $}    \\[1ex]
\hline
$n_{s } $ &  $0.9651^{+0.0049+0.0096}_{-0.0049-0.0093}  $  & $0.9741^{+0.0045+0.0088}_{-0.0045-0.0085} $    &  $0.9661^{+0.0045+0.0090}_{-0.0045-0.0089}  $  &  $0.9645^{+0.0045+0.0089}_{-0.0045-0.0088}   $   \\[1ex]

&  \textcolor{blue}{$0.9647^{+0.0049+0.0099}_{-0.0049-0.0094}    $}  & \textcolor{blue}{$0.9680^{+0.0040+0.0078}_{-0.0040-0.0079} $}  &  \textcolor{blue}{$0.9684^{+0.0047+0.0090}_{-0.0047-0.0090}    $}  &  \textcolor{blue}{$0.9701^{+0.0040+0.0080}_{-0.0040-0.0076} $}  \\[1ex]
\hline
$\tau_{\rm reio } $  & $0.076^{+ 0.015+0.030}_{-0.015-0.029} $   &  $0.072^{+0.013+0.025}_{-0.013-0.027}   $ & $0.080^{+ 0.014+0.028}_{-0.014-0.027} $     &  $0.073^{+0.014+0.027}_{-0.014-0.027} $   \\[1ex]

& \textcolor{blue}{$0.066^{+ 0.014+0.026}_{-0.014-0.028}$}   &  \textcolor{blue}{$0.073^{+0.012+0.023}_{-0.012-0.024}   $} & \textcolor{blue}{$0.075^{+ 0.013+0.026}_{-0.013-0.027}$}  &  \textcolor{blue}{$0.079^{+0.012+0.024}_{-0.012-0.023}   $} \\[1ex]
\hline
 $w_{\rm{dm0}}\,(95\% \, \rm CL) $  & $ <2.78\times 10^{-3}$ &  $ <  1.43\times 10^{-3}$   &  $ < 2.95\times 10^{-3}$ &   $ < 1.94\times 10^{-3} $ \\ [1ex]


 $w_{\rm{dm1}}\,(95\% \, \rm CL) $  & $ < 2.26\times 10^{-3}$ &  $ <  1.44\times 10^{-3}$  &  $ < 3.15\times 10^{-3}$ &   $ <  1.68\times 10^{-3}  $ \\ [1ex]


 $ \hat{c}^2_{\rm s,dm}\,(95\% \,\rm CL)$  &  $ <  2.18 \times 10^{-6} $  &  $ <  1.79 \times 10^{-6}$  &  $ <  2.31 \times 10^{-6}$ &   $<  1.95 \times 10^{-6}$ \\ [1ex]


\hline

$\Omega_{\rm{m0} }$   &   $0.279^{+ 0.022+0.036}_{-0.016-0.038}   $  & $0.292^{+0.008+0.017}_{-0.008-0.016}   $  &   $0.264^{+ 0.014+0.028}_{-0.014-0.026}   $  &   $0.284^{+ 0.007+0.015}_{-0.007-0.014}   $  \\[1ex]

& \textcolor{blue}{$0.312^{+0.009+0.017}_{-0.009-0.017}   $}  & \textcolor{blue}{$0.304^{+0.006+0.012}_{-0.006-0.012}   $} & \textcolor{blue}{$0.303^{+0.007+0.016}_{-0.008-0.014}   $}  &  \textcolor{blue}{$0.300^{+ 0.006+0.012}_{-0.006-0.012}   $}   \\[1ex]
\hline
 $\Omega_{\rm{\Lambda} }$ & $0.721^{+0.016+0.038}_{-0.022-0.035} $   & $0.707^{+0.008+0.016}_{-0.008-0.017} $  & $0.736^{+0.014+0.026}_{-0.014-0.028} $   &   $0.715^{+ 0.008+0.015}_{-0.008-0.015}  $  \\[1ex]
 
 & \textcolor{blue}{$0.688^{+0.009+0.017}_{-0.009-0.017} $}  & \textcolor{blue}{$0.695^{+0.006+0.012}_{-0.006-0.012} $}  & \textcolor{blue}{$0.697^{+0.008+0.014}_{-0.007-0.016} $}    &  \textcolor{blue}{$0.700^{+0.006+0.012}_{-0.006-0.012} $ } \\[1ex]
 \hline 
  $H_{\rm 0}$ &  $70.50^{+ 1.40+3.60}_{-2.10-3.20}    $  & $69.26^{+0.73+1.50}_{-0.73-1.40} $
 &  $72.00^{+ 1.40+2.70}_{-1.40-2.70}    $ &  $69.93^{+ 0.71+1.40}_{-0.71-1.30}  $\\[1ex]
 
 &  \textcolor{blue}{$67.53^{+0.64+1.30}_{-0.64-1.30} $}  & \textcolor{blue}{$68.08^{+0.47+0.91}_{-0.47-0.90} $} &  \textcolor{blue}{$68.18^{+0.59+1.10}_{-0.59-1.20} $}  &  \textcolor{blue}{$ 68.45^{+ 0.46+0.92}_{-0.46-0.91}  $} \\[1ex]
 \hline
  $\sigma_{8}$ &  $0.749^{+ 0.093+0.130}_{-0.050-0.160}  $ & $0.749^{+0.085+0.110}_{-0.040-0.140}  $ &  $0.747^{+ 0.110+0.130}_{-0.054-0.170}  $ &  $0.745^{+ 0.091+0.120}_{-0.049-0.140}  $ \\[1ex]
  
  & \textcolor{blue}{$ 0.817^{+0.009+0.017}_{-0.009-0.017} $}  & \textcolor{blue}{$0.819^{+0.009+0.018}_{-0.009-0.017}  $} & \textcolor{blue}{$ 0.819^{+0.009+0.017}_{-0.009-0.018} $}  &   \textcolor{blue}{$0.820^{+ 0.008+0.017}_{-0.008-0.016}$} \\[1ex]
 
  
  \hline \hline
\end{tabular}}
\end{table*}

\begin{figure*}
\includegraphics[width=17.cm]{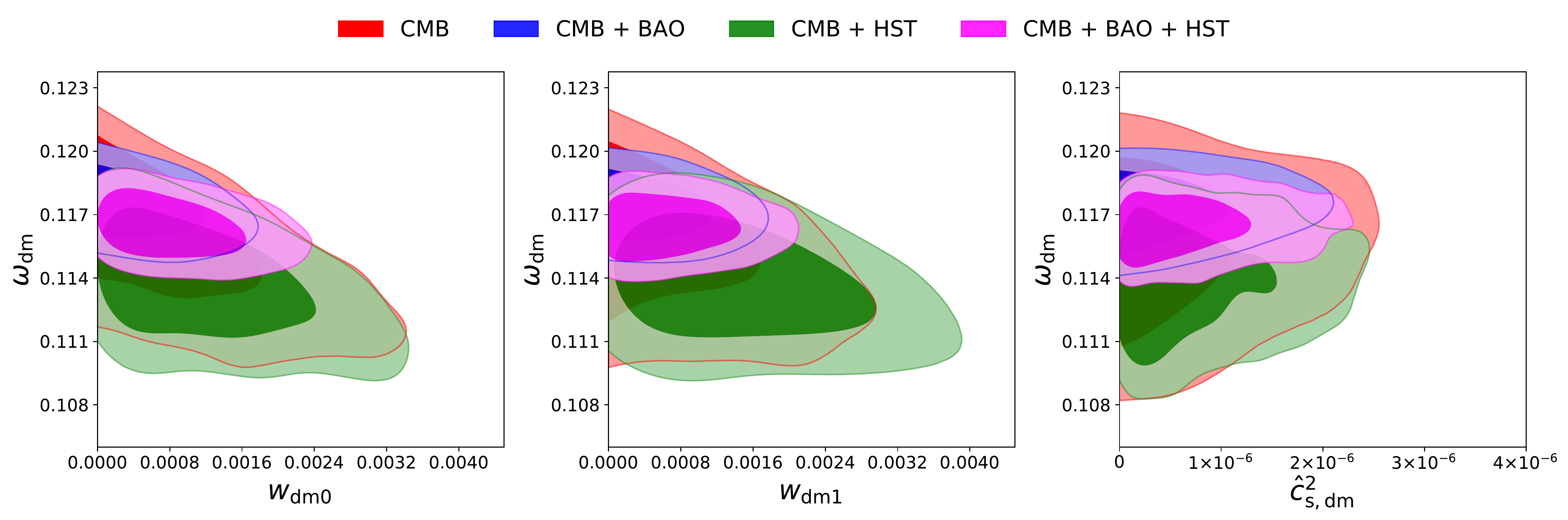} 
\caption{Two-dimensional marginalized distributions (68\% and 95\% CL) of some free parameters of $\Lambda$WDM model.} 
\label{fig1}
 \end{figure*}

We have implemented  the $\Lambda$WDM model in the publicly available CLASS \citep{blas2011cosmic} code, and used the Metropolis-Hastings algorithm in the parameter inference Monte Python \citep{audren2013conservative} code with uniform priors (as displayed in Table \ref{tab:priors}) on the model parameters to obtain correlated  Monte Carlo Markov Chain samples. We have chosen $w_{\rm dm0} \geq 0$ and $w_{\rm dm1} \geq 0$ in this work, though in the literature some authors have presented constraints allowing negative range of the EoS parameter $w_{\rm dm}$ of DM, but they do not find negative $w_{\rm dm}$ suitable for well-motivated physics. For instance, in \cite{muller2005cosmological}, it is stated that there is no particle motivation for negative $w_{\rm dm}$. On the other hand, in \cite{kumar2014observational}, the constraints on $w_{\rm dm}$ are presented by choosing its positive prior range due to the possible degeneracy with dark energy at the background level. Also, it is demonstrated in \cite{barboza2015thermodynamic} that all physical species (baryons, photons, neutrinos, dark matter etc.) must satisfy certain conditions on their EoS, in order to be stable from thermodynamics point of view, and here the DM fluid satisfies those conditions naturally with $w_{\rm dm0} \geq 0$ and $w_{\rm dm1} \geq 0$.  In the present work, the observational constraints on all model parameters are obtained by using four different data combinations: CMB, CMB + BAO, CMB + HST, and  CMB + BAO + HST.  The convergence of the Monte Carlo Markov Chains has been ensured by Gelman-Rubin criterion \citep{gelman1992inference}, which requires $1-R$ should be less than $0.01$ for all the parameters, in general. We have used the GetDist Python package to analyze the samples.



\section{Results and Discussion}
\label{results}
Table \ref{Table_M2} summarizes the observational constraints on the parameters  of the  $\Lambda$WDM model with four combinations of the data sets: CMB, CMB + BAO, CMB + HST, and  CMB + BAO + HST. The corresponding constraints on the $\Lambda$CDM parameters are displayed (in blue color) for comparison purpose. The constraints on the three extended DM parameters are given with upper bounds at 95\% CL. As expected, we see very tight constraints on these parameters of DM: the constraints on both the EoS parameters $w_{\rm{dm0}}$ and $w_{\rm{dm1}}$ of DM are of order $10^{-3}$ at 95\% CL, and  the constraint on the sound speed $\hat{c}^2_{\rm s,dm}$ of DM is of the order $10^{-6}$ at 95\% CL, from all the four data combinations. We note that the most tight constraints  are imposed by CMB + BAO data where the three parameters $w_{\rm dm0}$, $w_{\rm dm1}$  and $\hat{c}^2_{\rm s,dm}$ are respectively constrained to be less than $1.43\times 10^{-3}$, $1.44\times 10^{-3}$ and $1.79\times 10^{-6}$ at 95\% CL. From all the data combinations, we find that the constraints on all the three extended parameters of DM are consistent with zero at 95\% CL. This shows that CDM paradigm is consistent with the present observational data used in this study.   However, there are some interesting consequences on the standard $\Lambda$CDM dynamics via the small corrections of the extended DM parameters within their observed bounds, even if there is no enough statistical evidence to deviate from the CDM paradigm, as we will see in the following. 

\begin{figure}
\includegraphics[width=8cm]{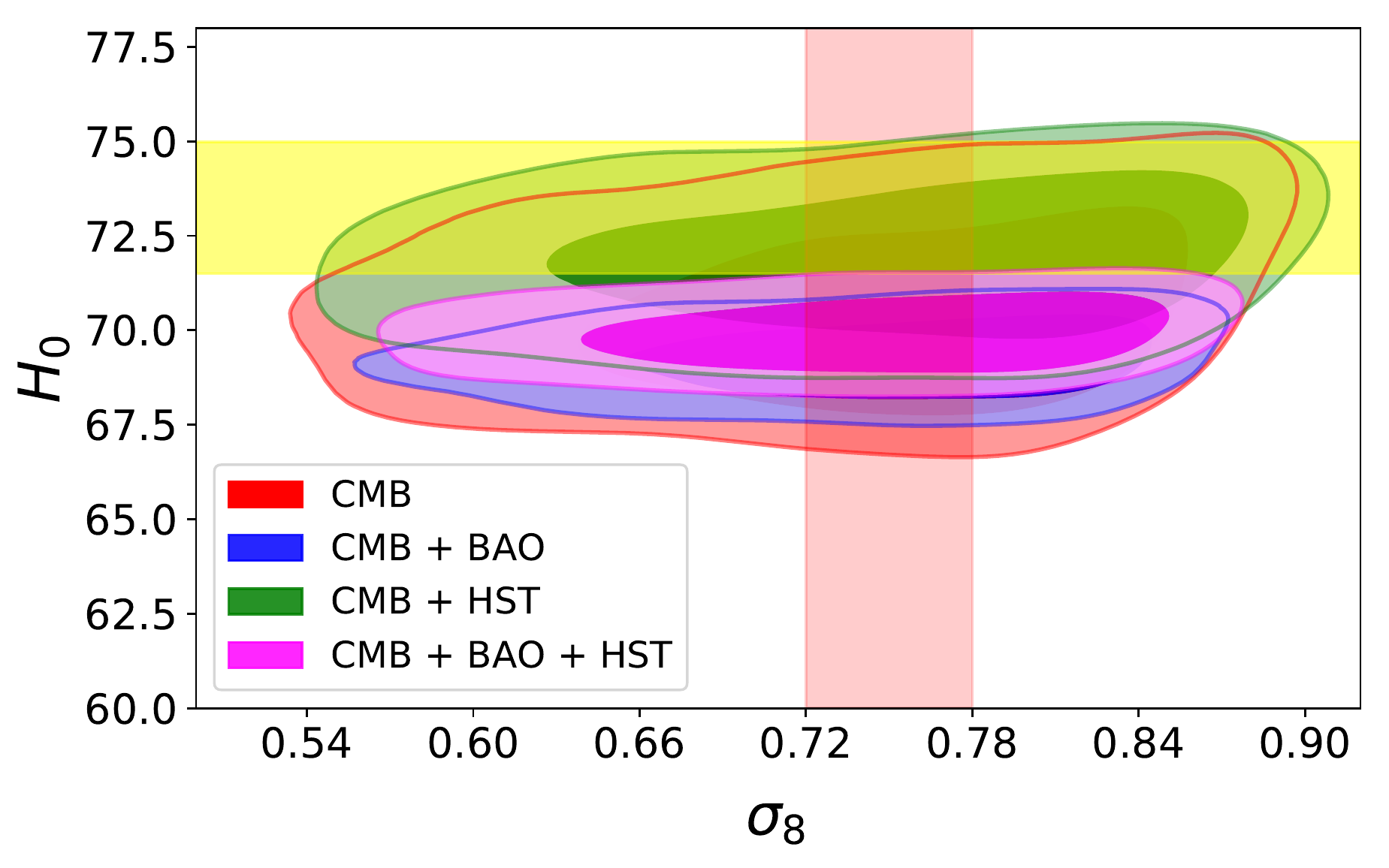}  
\caption{Parametric space in the plane $H_0 - \sigma_8$ for $\Lambda$WDM  from four different data combinations. The horizontal yellow band corresponds to $H_0 = 73.24 \pm 1.74$ km s$^{-1}$ Mpc$^{-1}$ \citep{riess20162} whereas the vertical light red band corresponds to $\sigma_8 = 0.75 \pm 0.03$ \citep{ade2014planck}.}
\label{fig_h0s8}
 \end{figure}

\begin{figure*}
     \centering
     \includegraphics[width=8.8cm]{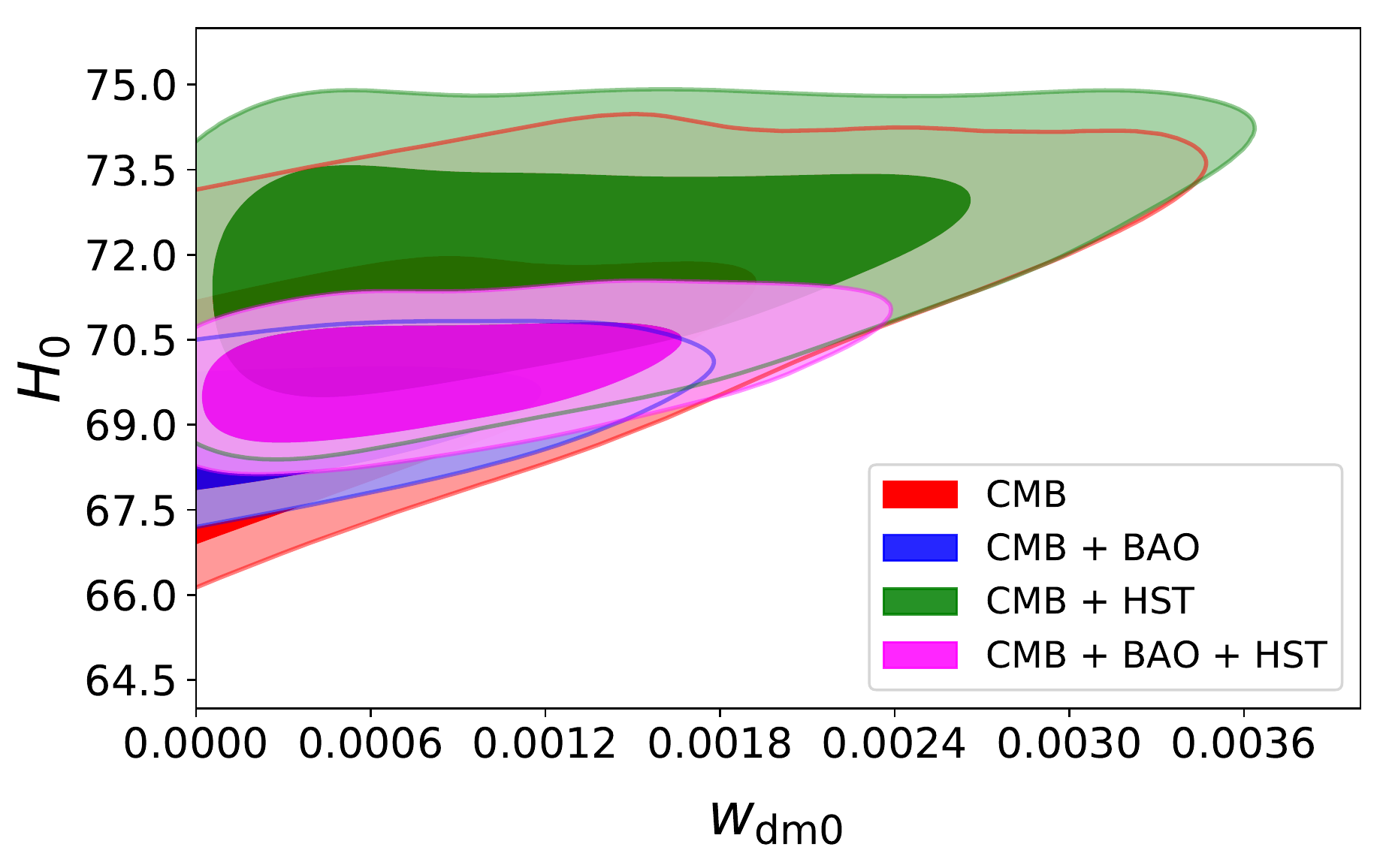}
     \includegraphics[width=8.8cm]{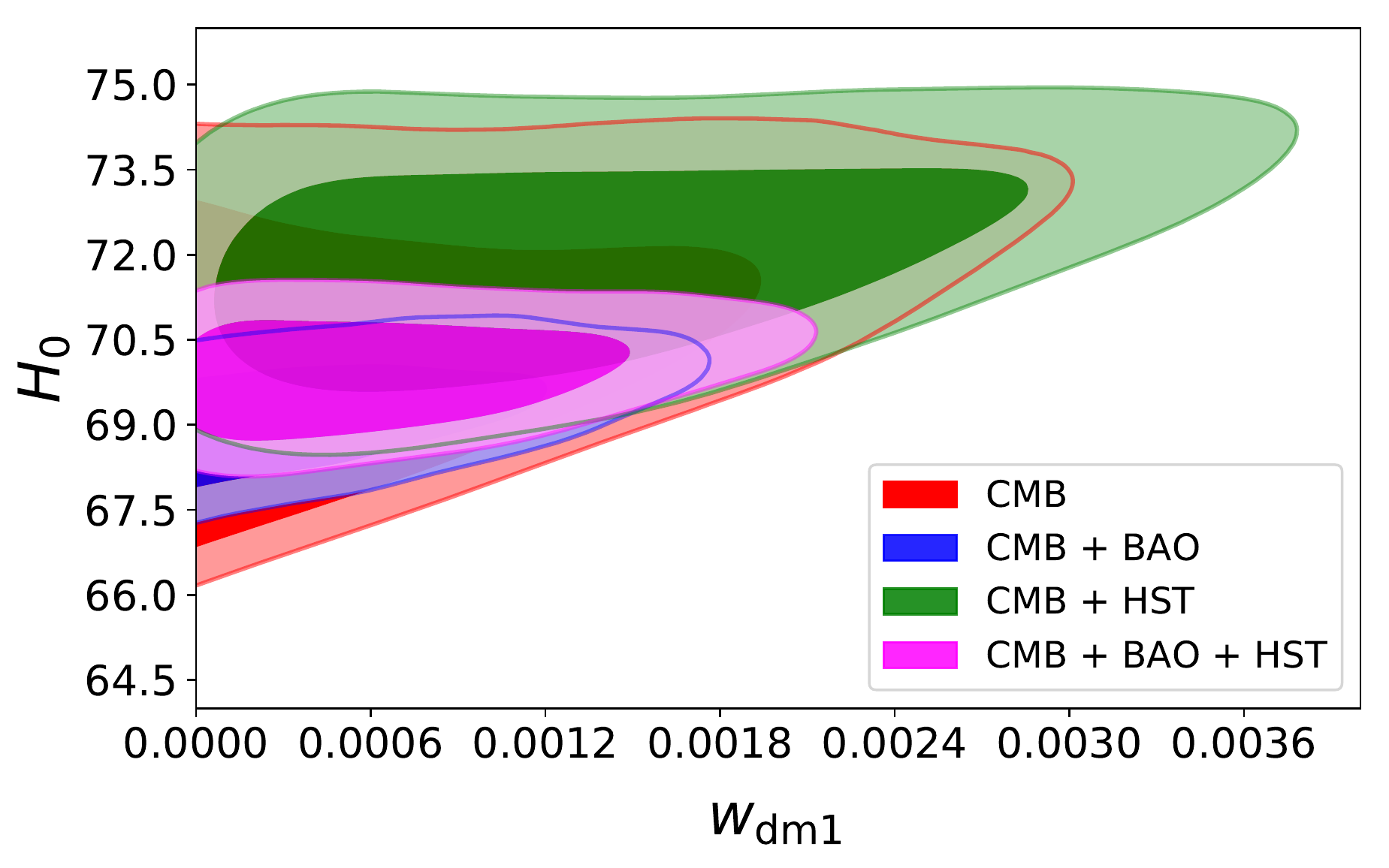}
     \caption{Two-dimensional marginalized distributions (68\% and 95\% CL) of  $H_0$ vs  EoS  parameters, $w_{\rm dm0}$ and $w_{\rm dm1}$ of DM for $\Lambda$WDM model.}
     \label{fig_wh0}
 \end{figure*}

From Figure \ref{fig1}, we observe a small negative correlation of the DM EoS parameters $w_{\rm{dm0}}$ and $w_{\rm{dm1}}$ with $\omega_{\rm dm }$. It implies the larger values of $w_{\rm{dm0}}$ and $w_{\rm{dm1}}$ would correspond to smaller values of $\omega_{\rm dm }$. Consequently in Table \ref{Table_M2}, we see smaller mean values of $\omega_{\rm dm }$ in comparison to the $\Lambda$CDM model, in all four cases of data combinations. Similarly, we notice smaller mean values of $\omega_{\rm b}$ in all cases. Consequently, we find smaller mean values of the derived parameter $\Omega_{\rm m }$ and larger mean values of $\Omega_{\Lambda }$ in comparison to the $\Lambda$CDM model (see Table \ref{Table_M2}). The derived parameters $H_0$ and $\sigma_8$, representing the present Hubble expansion rate of the Universe and amplitude of present matter density fluctuation in a sphere of the radius of $8h^{-1}$Mpc, respectively, are also affected significantly due to the inception of the extended DM parameters. It can be seen from Table \ref{Table_M2} that the variability of EoS of DM provides the higher mean values of Hubble constant (as compared to $\Lambda$CDM). We have $H_0 = 70.50^{+1.40}_{-2.10}$ km s$^{-1}{}$Mpc$^{-1}$ at 68\% CL from Planck CMB data alone. The inclusion of BAO data yields slightly lower mean value,  $ H_0= 69.26 \pm 0.73$ km s$^{-1}{}$Mpc$^{-1}$ at 68\% CL with significantly small errors which are Gaussian in nature. It is worthy to mention that due to the less DM abundance (effect of varying DM EoS) as compared to the $\Lambda$CDM model, we have higher mean values of Hubble constant even without using HST prior. The constraints presented here on $H_0$ from CMB and CMB + BAO data combinations are stronger than the constraints obtained in a similar analysis by \cite{thomas2016constraining} with the same data combinations, where a constant EoS of DM was assumed. The inclusion of HST prior in analysis significantly improves the constraints to $H_0 = 72.00 \pm 1.40$ km s$^{-1}{}$Mpc$^{-1}$ at 68\% CL, favoring locally measured value of Hubble constant. The constraint with the combined analysis: CMB + BAO + HST, gives $ H_0= 69.93 \pm 0.71$ km s$^{-1}{}$Mpc$^{-1}$ at 68\% CL which is almost same as with CMB + BAO combination. Also, see Figure  \ref{fig_h0s8}, which shows the parametric space in the plane $H_0$ - $\sigma_8$ for $\Lambda$WDM model from the four  data combinations. We see that the confidence region for the combination CMB + BAO + HST almost overlaps with the region from CMB + BAO with a little shift in the mean value of $H_0$ to the higher side. Thus, the $\Lambda$WDM model mildly favors the value of Hubble constant from the local measurement. We also observe that the parameter $H_0$ is positively correlated with both the EoS parameters of DM as may be noticed from Figure \ref{fig_wh0}.

 \begin{figure}
    \centering
    \includegraphics[width=8cm]{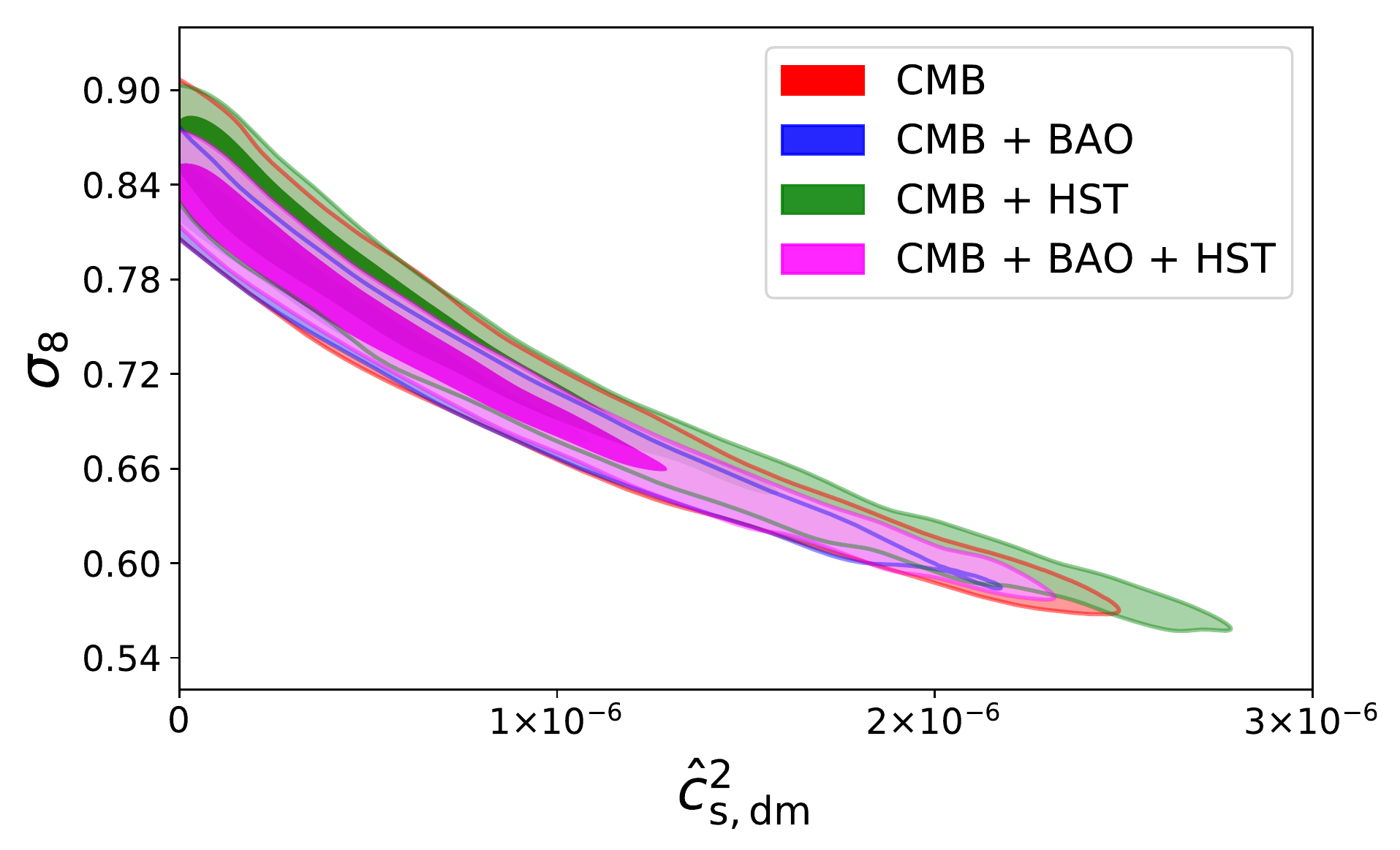}
    \caption{Parametric space in the plane $\hat{c}^2_{\rm s,dm}-\sigma_8$ with all data combinations.}
    \label{cs2-s8}
\end{figure}
 The sound speed of DM has a strong degeneracy with the derived parameter $\sigma_8$. This is due to the fact that this parameter sufficiently reduces the growth of matter density fluctuations on the length scales below the diffusion length scale \citep{thomas2016constraining}. The amplitude of this matter density fluctuation is characterized by $\sigma_8$, resulting in a strong degeneracy between this parameter and $\hat{c}^2_{\rm s,dm}$. Figure \ref{cs2-s8} represents the parametric space in the plane $\hat{c}^2_{\rm s,dm}$ - $\sigma_8$ with all the four data combinations. We can see that $\hat{c}^2_{\rm s,dm}$ is negatively correlated with $\sigma_8$.
 In Table \ref{Table_M2}, we see lower mean values of $\sigma_8$ with all the data combinations but with large errors in each case (compared to the $\Lambda$CDM model). These large errors are due to the strong degeneracy between $\hat{c}^2_{\rm s,dm}$ and $\sigma_8$. Thus, the presence of the sound speed of DM provides significantly lower mean values of $\sigma_8$ consistent with LSS observations. One can also see from Figure \ref{fig_h0s8} that the vertical red band, representing the range of $\sigma_8$ measured by LSS observations \citep{ade2014planck}, passes through the central region of each contour.

From Table \ref{Table_M2}, we notice in general that the inclusion of BAO data significantly tightens the constraints on model parameters whereas the addition of HST prior does not do so. The addition of HST prior to CMB data yields higher mean value of $H_0$ consistent with the local measurement in the $\Lambda$WDM model but not in $\Lambda$CDM model. Also, we observe higher mean values of $H_0$ with other three data combinations in comparison to the $\Lambda$CDM. Thus, the underlying $\Lambda$WDM model equipped with significant positive values of the extended DM parameters might reconcile the $H_0$ tension.



 \subsection{Estimating the warmness}

Without loss of generality, the warmness of DM particles can be estimated by its dynamic character determined by $w_{\rm dm}(a)$. Relaxing the condition $w_{\rm dm} \equiv 0$ and going beyond the non-relativistic limit, we can write  

\begin{equation}
w_{\rm dm} \equiv \frac{p_{\rm dm}}{\rho_{\rm dm}} \simeq \frac{T_{\rm dm}}{m_{\rm dm}},
\end{equation}
where $T_{\rm dm}$ is the DM temperature \citep{armendariz2014cold}. Assuming that DM particles interact with other species only gravitationally, we have $T_{\rm dm} = T_{\rm dm0}\;a^{-2}$, where $T_{\rm dm0}$ is the temperature of DM today. Further,

\begin{equation}
\label{DM_T}
T_{\rm dm0} = w_{\rm dm0}\; m_{\rm dm}.
\end{equation}
Thus, from our estimates of $w_{\rm dm0}$, and for a given $m_{\rm dm}$ scale, $T_{\rm dm0}$ can be easily calculated.

Within the minimal $\Lambda$CDM model, DM particles are assumed to be cold in the strict non-relativistic limit $T_{\rm dm}/m_{\rm dm} \longrightarrow 0 $. Thus, we can think a possible deviation from this limit as a test for the warmness of DM particles. On the other hand, the relativistic limit (for a possible hot species) is determined by $T_{\rm dm}/m_{\rm dm} \gg 1 $. Thus, the warm species must lie between these limits, and here we can quantify it by measuring $w_{\rm dm}(a)$. For any possible evidence of $w_{\rm dm}(a) \neq 0 $, we can relax the condition that DM is purely cold, with associated background temperature today given by (\ref{DM_T}), quantifying its warmness.

In order to quantify the warmness precisely, we need to determine $m_{\rm dm}$ as required in  (\ref{DM_T}). Following standard procedures, the fitting formula from Boltzmann code calculations for the free-streaming on matter distribution is given by a relative  transfer function \citep{bode2001halo}:

\begin{equation}
T_{\rm wdm} = \left[\frac{P_{\rm wdm}(k)}{P_{\rm cdm}(k)} \right]^{1/2} = [1 + (\alpha k)^{2.24} ]^{-4.46},
\end{equation}
where the parameter $\alpha$ is given by

\begin{equation}
\label{alpha_WDM}
\alpha = 0.049 \Big( \frac{m_{\rm dm}}{\rm k eV} \Big)^{-1.11} \Big(\frac{\Omega_{\rm cdm}}{0.25} \Big)^{0.11} \Big(\frac{h}{0.7} \Big)^{1.22} h^{-1} \rm Mpc.
\end{equation}

This fitting function  applies to the case of thermal relics, and we use it to estimate $m_{\rm dm}$ values. For example, let us choose $\alpha = 0.1$ $h^{-1} \rm Mpc$ \citep{viel2005constraining}, though, in general, $\alpha$ should be fit together with other free parameters of the model baseline during the MCMC analysis. Also, it may be noted that by using the constraint on $\alpha$ obtained in \cite{viel2005constraining}, we implicitly use an additional dataset of Lyman-alpha forest data. Certainly it can bias the results since possible larger $\alpha$ values can lead to smaller borders on $m_{\rm dm}$. But, here, we keep this $\alpha$ upper value for qualitative estimates. Also, we do not assume corrections on non-linear scale, where warm DM properties should manifest significantly, beyond the default modeling implemented in CLASS code. Thus, taking the above mentioned value of $\alpha$ to estimate $m_{\rm dm}$ seems reasonable for simple and qualitative lower bound estimate of $m_{\rm dm}$. In Table \ref{DM_mass}, we summarize the corresponding lower bound on DM mass for all considered data combinations. The estimates on $m_{\rm dm}$ are simply made by direct substitution of the best-fit mean values of the parameters from our analyzes in eq. (\ref{alpha_WDM}). 
\\

\begin{table}
\caption{Lower bounds on DM mass $m_{\rm dm}$ in the units of keV from four data combinations. } 
\label{DM_mass}
\begin{tabular}{l c  }
\hline \hline
Data  &  $\Lambda$WDM    \\
\hline
CMB               &   $0.526$   \\
CMB + BAO         &   $ 0.519$    \\
CMB + HST         &   $0.537$  \\
CMB + BAO + HST   &   $0.526$   \\

\hline  \hline
\end{tabular}
\end{table}

For all data combinations, we notice that $m_{\rm dm} > 0.5$ keV, thus compatible with the Tremaine-Gunn bound \citep{tremaine1979dynamical}, that allows structure formation. \cite{narayanan2000constraints} have constrained the mass of a thermal warm DM particle (assuming to account for all the DM content, like here) to be $m_{\rm dm} > 0.75$ keV. We see that we have also deduced the $m_{\rm dm}$ values in the same order of magnitude. Other borders on warm DM are discussed in \cite{colombi1995large,fabris2012testing,inoue2015constraints,gariazzo2017cosmological,murgia2017non,lopez2017warm,schneider2018constraining,hipolito2018general,martins2018forecasts}. Now, an estimate on the warmness of DM can easily be obtained by evaluating the DM temperature today, $T_{\rm dm0}$, using (\ref{DM_T}), and also the evolution DM temperature with the cosmic time, given by $T_{\rm dm}(a) = T_{\rm dm0}\; a^{-2}$.

\subsection{Comparison with previous studies}
 
The DM EoS parameter has been constrained by many authors in the literature as mentioned in the introduction.  \\
 Constant EoS of DM has been constrained by \cite{muller2005cosmological} using CMB, SNe Ia,  and LSS data: $w_{\rm dm0}<{\mathcal{O}(10^{-3})}$ at 99.7\% CL, assuming vanishing adiabatic sound speed. \cite{calabrese2009cosmological} have placed the constraints:  $w_{\rm dm0} <{\mathcal{O}(10^{-2})} $ at 95\% CL from WMAP alone, which is weaker constraint by an order of magnitude on $w_{\rm dm0}$ at 95\% CL found in the present  work. 
In addition, they have found that combining CMB data with SNe Ia, SDSS and HST improves the constraint to $w_{\rm dm0} <{\mathcal{O}(10^{-3})} $ at 95\% CL.  \cite{xu2013equation} have  constrained the EoS of DM by  using Planck 2013, BAO, SNe Ia and found $w_{\rm dm0} <{\mathcal{O}(10^{-3})} $ at 99.7\% CL. They have also examined the effect of WiggleZ measurement of the matter power spectrum, and found that it has a small effect on $w_{\rm dm0}$ with bound still of the order $10^{-3}$ at 99.7\% CL. In \cite{thomas2016constraining}, the generalized  DM parameters: the  EoS, sound speed, and viscosity ($c^2_{\rm vis,dm}$) (all are taken as constants) have been constrained by using the data from Planck-2015 together with BAO and HST. They have found the constraints:  $w_{\rm dm0} <{\mathcal{O}(10^{-3})} $  and  $\hat{c}^2_{\rm s,dm}$,  $c^2_{\rm vis,dm} <{\mathcal{O}(10^{-6})}$, all at 99.7\% CL. In the present work, we have observed that allowing a variable EoS DM provides significantly tighter upper bounds on parameter $w_{\rm dm0}$  than those found in \cite{thomas2016constraining}, where a constant EoS of DM was assumed. For instance, the tightest upper bound found in \cite{thomas2016constraining} is $w_{\rm dm0}<2.38 \times 10^{-3}$ at 99.7\% CL with the combination: CMB + BAO. In our case, the tightest upper bound is  $w_{\rm dm0}<1.80 \times 10^{-3}$ at 99\% CL with same data combination: CMB + BAO. Similarly, with other data combinations, significantly tighter upper bounds are found on $w_{\rm dm0}$ at 95\% CL (also 99\% CL) in comparison to the ones found by \cite{thomas2016constraining}. \cite{kunz2016constraints} have also found similar constraints on extended DM parameters (assuming all constants) by using the data from Planck including polarization with  geometric probes from SNe Ia  and BAO.
The constraints in both the above-mentioned works are in good agreement with our constraints on the extended DM parameters. It is important to mention that the present work differs from both  \cite{thomas2016constraining} and \cite{kunz2016constraints} in the fact that they have considered constant EoS of DM together with non-zero viscosity. In the present work, we have considered a time-varying EoS and zero viscosity of DM. Recently, \cite{tutusaus2018generalized} have constrained a model with constant EoS and  sound speed of DM with zero viscosity and found: $w_{\rm dm0} <{\mathcal{O}(10^{-3})} $  and $\hat{c}^2_{\rm s,dm} <{\mathcal{O}(10^{-6})} $, both at 68\% CL by using the data combination: CMB + SNIa + BAO. In addition, they have also shown that the photometric Euclid survey placed nice constraints on all parameters, in particular, a very strong constraint on the sound speed of DM.
The EoS of DM has recently been constrained in \cite{kopp2018dark} by allowing it to vary in eight redshift bins from $z=10^5$ to present time ($z=0$), assuming sound speed and viscosity equal to zero, and found that EoS of DM does not deviate significantly from the null value at any time.

In short, as expected, there are small corrections on the extended DM parameters $w_{\rm dm0}$, $w_{\rm dm1}$ and $\hat{c}^2_{\rm s,dm}$ in our results in line with the literature. The temporal dependence of $w_{\rm dm}$ in our work is quantified by $w_{\rm dm1}$, and our analysis is a null test of the $w_{\rm dm} = {\rm constant}$ case in the literature via the CPL parametrization of $w_{\rm dm}$. We notice that $w_{\rm dm0}$ and $w_{\rm dm1}$ are equally preferred/constrained by the considered data in the order of magnitude. In other words, the presence of $w_{\rm dm1}$ is not neglected by the data in comparison to the $w_{\rm dm} = {\rm constant}$ case. The parameter $w_{\rm dm1}$ shows correlation with other parameters similar to $w_{\rm dm0}$, as may be seen in Figure \ref{fig1} and Figure \ref{fig_wh0}. The presence of $w_{\rm dm1}$ minimally relaxes the constraints on the full model baseline parametric space in comparison to the $w_{\rm dm} = {\rm constant}$ case in earlier studies.


 \begin{figure}
 \includegraphics[width=8.0cm]{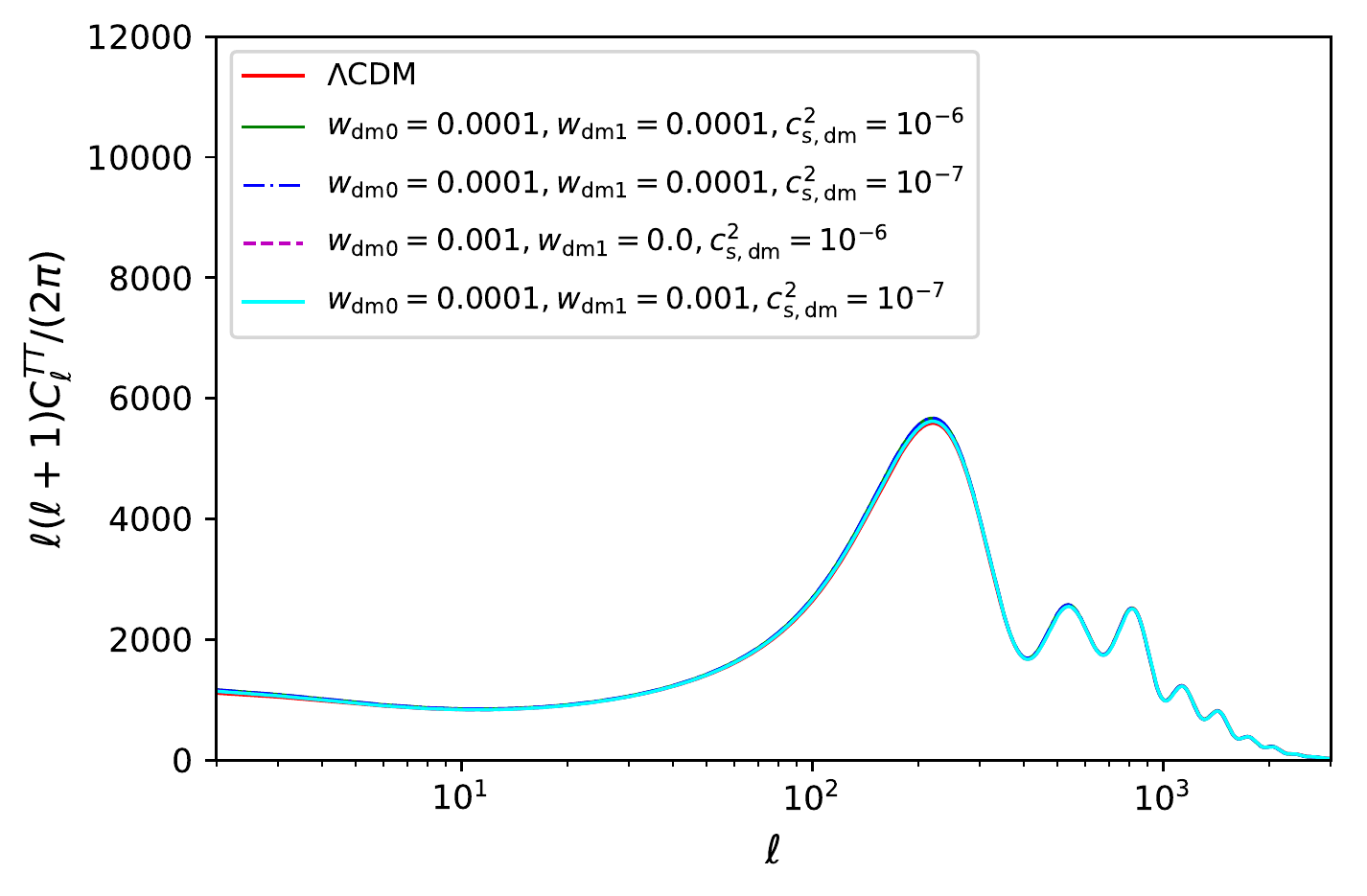}
 \caption{The CMB TT power spectra from base line Planck 2015 $\Lambda$CDM model for some values of model parameters as mentioned in the legend while other relevant parameters are fixed to their mean values as shown in Table \ref{Table_M2}.} 
 \label{fig_cmb}
  \end{figure}
 \begin{figure}
\includegraphics[width=8.0cm]{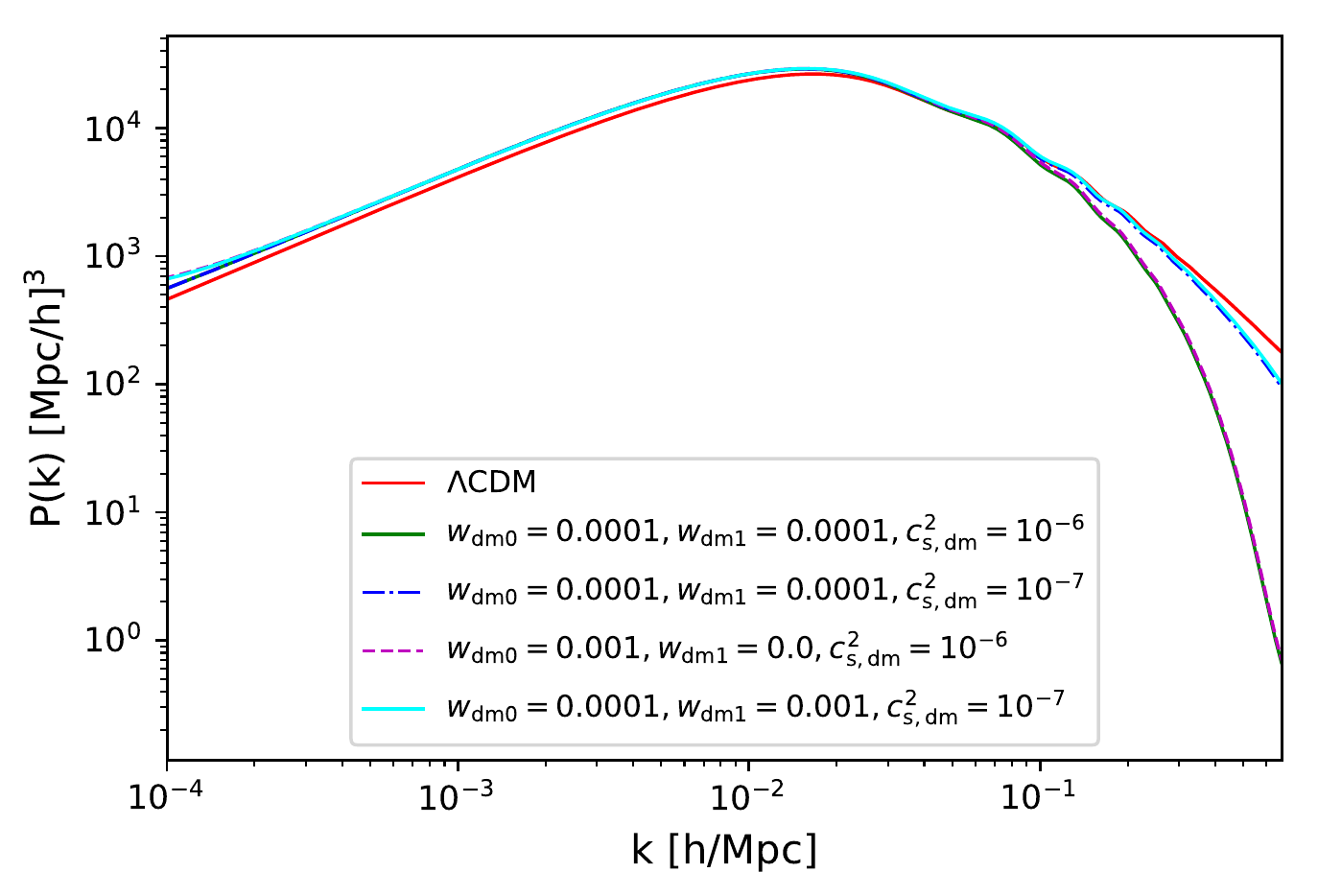} 
 \caption{The matter power spectra from base line Planck 2015 $\Lambda$CDM model for some values of model parameters as mentioned in the legend while other relevant parameters are fixed to their mean values as shown in Table \ref{Table_M2}.} 
  \label{fig_matter}
 \end{figure}

\subsection{Effects on CMB TT and matter power spectra}
\label{powerspectra}

 In this study, we have considered the possibility of $w_{\rm dm} > 0$. It should generate effects on CMB TT similar to the ones arising due to the change in $\omega_{\rm dm}$ (dimensionless DM density), modifying the heights of the first few acoustic peaks. In fact, $w_{\rm dm} > 0$ should increase the DM density at the time of radiation-matter equality and, therefore modify the modes that enter the horizon during radiation domination, leading to a suppression around the acoustic peak scales. Also, the acoustic peaks in the CMB depend on the angular diameter distance to the last
scattering, where it is influenced by the change in the expansion rate $H$, which is modified here by the presence of the dynamical $w_{\rm dm}$ term. As $w_{\rm dm}$ increases, the angular diameter distance to the last scattering surface decreases such that features are shifted to smaller angular scales. Larger values for $w_{\rm dm}$ would result in behaving more like radiation for DM, generating large acoustic driving and boosting on the CMB peaks, with the prevision that should be inconsistent with observations. Also, the major indirect evidence for DM comes from the CMB peaks. The possible absence of CDM particles would introduce large acoustic driving,  boosting the peak amplitude, which can also lead to a spectrum that completely disagrees with observations. That is why data from CMB, in general, leads to very strong limits like $w_{\rm dm} \ll 1$. The effective sound speed $\hat{c}^2_{\rm s, dm}$ parameter will influence the spectrum on acoustic peak similar to  $w_{\rm dm}$, where $\hat{c}^2_{\rm s, dm} > 0$ would cause the amplitude of the acoustic peaks to decrease relative to the large scale anisotropy. At large scales, where the integrated Sachs-Wolfe effect is predominant, the main effect of $\hat{c}^2_{\rm s, \rm dm} > 0$ is to increase the gravitational potential decay after recombination until the present time, causing an increase to the anisotropy for $l < 40$. Changing $w_{\rm dm} > 0$ has a very mild effect on the integrated Sachs-Wolfe, compared to  $\hat{c}^2_{\rm s, dm}$, but both parameters with positive variation yield the same effects on large scales.
Figure \ref{fig_cmb} shows how the parameters, $w_{\rm dm0}$, $w_{\rm dm1}$ and $\hat{c}^2_{s,\rm dm}$ can affect the CMB TT spectrum. We note small and significant deviations on the minimal $\Lambda$CDM model as described above, where the effects on the acoustic peaks (i.e., effects for $l > 50$) are less noticeable. Possible effects to shift the spectrum into the direction to smaller angular scales are minimal due to very small corrections (insignificant corrections) on the angular diameter distance at last scattering as the effects of $w_{\rm dm0}$, $w_{\rm dm1} \ll 1$ are very small. 


On the other hand, the LSS of the Universe also depends directly on the DM properties. As pointed out in \cite{hu1998structure}, the clustering scale becomes independent of the DM EoS,  and DM extended properties should change only the amplitude of the perturbations, that can be observed by looking at the matter power spectrum. We also expect these changes basically to be in the order of magnitude compatible with the observed Universe. Therefore, we set the free parameters within the limits derived here up to 99\% CL. The presence of $\hat{c}^2_{\rm s, dm} > 0$ decreases the amplitude on $P(k)$, and in return $w_{\rm dm0} > 0$ increases the amplitude of perturbations. Figure \ref{fig_matter} shows $P(k, z =0)$ for some selected values of $\hat{c}^2_{\rm s,dm}$, $w_{\rm dm0}$ and $w_{\rm dm1}$. In general, we notice that $w_{\rm dm0}$ influences more the amplitude than $\hat{c}^2_{\rm s,dm}$, causing a net increase in the amplitude of the matter power spectrum.


\subsection{Bayesian model comparison}

 In the present work, we have analyzed an extension of the standard $\Lambda$CDM model. Thus, apart from parameter estimation performed here, it is important to perform a statistical comparison of the considered model with a  well-fitted standard model (reference model). For this purpose, we use the Akaike Information Criteria (AIC) \citep{akaike1974anew, anderson2004model}, defined as 
 
 \begin{equation} \nonumber
 \text{AIC} = -2 \ln  \mathcal{L}_{\rm max} + 2N \quad = \chi_{\rm min}^2 + 2N,
 \end{equation}
 where $ \mathcal{L}_{\rm max}$ is the maximum likelihood function of the model, and $N$ is the total number of free parameters in the model baseline. To compare a model $i$ under consideration with a reference model $j$ (well-known best-fit model), we need to calculate the AIC difference between two models, i.e.,  $\Delta{\rm AIC}_{ij} =  \text{AIC}_{i}- \text{AIC}_{j}$. This difference can be interpreted as the evidence in favor of the model $i$ compared to the model $j$. As argued in \cite{tan2012reliability}, one can assert that one model is better than the other if the AIC difference between the two models is greater than a threshold value $\Delta_{\rm threshold}$. According to the thumb rule of AIC, $\Delta_{\rm threshold} = 5$  is a universal value of threshold regardless of the properties of the model considered for comparison. It has been mentioned in \cite{liddle2007information} that this threshold is the minimum AIC difference between two models to strongly claim that one model is better compared to the other model. Thus, an AIC difference of 5 or more favors the model with smaller AIC value.

\begin{table}
\caption{\label{evidence}{}Difference of AIC values  of $\Lambda$WDM model under consideration with respect to $\Lambda$CDM model (reference model) with four data combinations. }
\begin{tabular}{l c  }
\hline \hline
Data  &  $\Delta \rm AIC_{\Lambda WDM}$  \\
\hline
CMB        &        7.78     \\
CMB + BAO  &         5.52   \\
CMB + HST   &     2.62    \\
CMB + BAO + HST    &     2.10     \\

\hline \hline
\end{tabular}
 \end{table}

Table \ref{evidence} summarizes the $\Delta \rm AIC$ values of the considered model for all the data combinations. We have $\Delta \rm AIC$ values greater than the threshold value for the data: CMB and CMB + BAO. Therefore, it can be claimed that the standard $\Lambda$CDM model is strongly favored over the $\Lambda$WDM model with data combinations: CMB and CMB + BAO. On another hand, for CMB + HST and CMB + BAO + HST combinations, we can not claim statistical evidence in favor of either of models since $\Delta \rm AIC$ values are much less than the threshold value.
 \section{Concluding remarks}

The presence of DM plays a crucial role in explaining the current cosmological data wherein it is almost impossible to explain the origin of CMB and LSS without the presence of this dark component. Despite being a key ingredient in modern cosmology, the nature of DM is one of the most open questions in contemporary science, and its general properties like spin, mass, interaction cross section, lifetime, etc, are not yet completely closed for study via phenomenological attempts. In the present work, we have investigated an extension of the $\Lambda$CDM model via the extended properties of DM:  a possible time dependence of EoS of DM via the CPL parametrization $w_{\rm dm} = w_{\rm dm 0} + w_{\rm dm 1}(1-a)$, and the non-null sound speed $\hat{c}^2_{\rm s,dm}$. Analyzing these properties by the data summarized in section \ref{data}, we have derived new and robust constraints on the extended free parameters of DM. The most tight constraints  are imposed by CMB + BAO data where the three parameters $w_{\rm dm0}$, $w_{\rm dm1}$  and $\hat{c}^2_{\rm s,dm}$ are respectively constrained to be less than $1.43\times 10^{-3}$, $1.44\times 10^{-3}$ and $1.79\times 10^{-6}$ at 95\% CL (see Table \ref{Table_M2}), which are in line with the results in the literature. Thus, the extended parameters of DM are strongly constrained, and all show consistency with zero at 95\% CL, indicating no evidence beyond the CDM paradigm. Further, the extended properties of DM significantly affect several parameters of the base $\Lambda$CDM model. In particular, in all the analyses performed here, we have found significantly larger mean values of $H_0$ and lower mean values of $\sigma_8$ in comparison to the base $\Lambda$CDM model. Thus, the well-known $H_0$ and $\sigma_8$ tensions might be reconciled in the presence of extended DM parameters within the $\Lambda$CDM framework. Also, we estimate the warmness of DM particles as well as its mass scale, and find a lower bound: $\sim$ 500 eV from our analyses, compatible with the Tremaine-Gunn bound and other such limits found in the literature. Here, it deserves mention that we have given some qualitative estimates of the DM mass scale from the results of our analyzes. It could be worthwhile to investigate our model for direct and precise constraints on the DM mass scale using the approach followed by \cite{viel2005constraining}.

From our analyzes, it is clear that even a little deviation provided by the extended DM properties could lead to interesting, useful and significant changes in the evolution of the base $\Lambda$CDM Universe. So it would be worthwhile to investigate extended DM properties in the light of forthcoming data from various surveys/experiments in near future. As argued in \cite{hawking1966perturbations}, and recently extended and well-determined in \cite{flauger2018gravitational}, the presence of a medium with non-zero shear viscosity, can lead the propagation of the gravitational waves to dissipation due to damping effect. It could be interesting to study DM extended properties that can induce possible effects of shear viscosity, and investigate its limits using gravitational wave physics.

\section*{Acknowledgments}
S.K. gratefully acknowledges the support from SERB-DST project No. EMR/2016/000258, and DST FIST project No.
SR/FST/MSI-090/2013(C). R.C.N would like to thank FAPESP for financial support under grant \# 2018/18036-5. S.K.Y. acknowledges the Council of Scientific \& Industrial Research (CSIR), Govt. of India, New Delhi, for awarding Senior Research Fellowship (File No. 09/719(0073)/2016-EMR-I). 

\bibliographystyle{mnras}
\bibliography{reference} 

\bsp	
\label{lastpage}
\end{document}